\documentstyle[12pt]{article}
\title{Spontaneous Baryogenesis in Supersymmetric Models}
\author{S.~A.~Abel$^{1}$, W.~N.~Cottingham$^{2}$ and
I.~B.~Whittingham$^{3}$\vspace{0.3cm}
\\$^{1}$Rutherford Appleton Laboratory
\\Chilton, Didcot
\\Oxon OX11 0QX
\\England
\\ \vspace{0.3cm}
\\$^{2}$H.~H.~Wills Physics Laboratory
\\Royal Fort, Tyndal Avenue
\\Bristol BS8 1TL
\\England
\\ \vspace{0.3cm}
\\$^{3}$Department of Physics
\\James Cook University
\\Townsville
\\Australia 4811}

\begin{document}

\small

\maketitle

\title{Abstract}

\begin{abstract}
In this paper we extend the results of previous work
on spontaneous baryogenesis to general models involving
CP violation in the Higgs sector. We
show how to deal with Chern-Simons terms appearing in the
effective potential arising from phase changes in the
vacuum expectation values of the Higgs fields. In
particular, this enables us to apply this mechanism
to general supersymmetric models including the
minimal supersymmetric standard model, and the extended
model with a gauge singlet.  A comparison
is made between this approach, and that in which one
solves the equations of motion for Higgs winding modes.
As anticipated in earlier work, the effect of the latter
approach is found to be small.

\end{abstract}

\pagebreak
\section{Introduction}
There has been considerable recent interest
\cite{CK88}-\cite{M92} in the possibility
of generating the observed matter-antimatter asymmetry of the universe
at the electroweak phase transition within the standard electroweak
model or its minimal extension to include an additional Higgs doublet
(for recent reviews see Ref.\cite{Dine92}).
It has been well known since the pioneering work of t'~Hooft \cite{HOOF}
that, due to the axial anomaly, baryon number $B$ and lepton
number $L$ are not conserved in the standard model. For $N_{g}$ fermion
generations the change in baryon and lepton number is given by
\begin{equation}
\label{BL}
\Delta B = \Delta L = N_{g} \, \Delta N_{CS}
\end{equation}
where $N_{CS}$, the Chern-Simons number for the gauge field, is given by
\begin{equation}
N_{CS} = \int d^{3}x \, K^{0}
\end{equation}
with $K^{\mu}$ the topological current
\begin{equation}
\label{topo}
K^{\mu} = \frac{g_2^{2}}{16 \pi^{2}} \varepsilon^{\mu\nu\rho\lambda} \,
Tr ( F_{\nu\rho} \, A_{\lambda} + \frac{2}{3}ig_2 A_{\nu}A_{\rho}A_{\lambda})
\end{equation}
and $A_{\mu}=A_{\mu}^{a} \,\sigma^{a}/2, F_{\mu\nu}=F_{\mu\nu}^{a} \,
\sigma^{a}/2$ the
$SU(2)$ gauge field and field strength respectively. In
Eq.(\ref{BL}) the
change in $N_{CS}$ is associated with transitions between the different
topological sectors of the gauge and Higgs fields. The different sectors
are separated by energy barriers corresponding to the saddle point
sphaleron configuration of mass $M_{sph} \approx 5M_{W}/\alpha_W$ (where
$\alpha_W=g^{2}_{2}/4\pi$)
so that these transitions are heavily suppressed at zero temperature
where $M_{sph} \approx
10TeV$ and the $\Delta N_{CS}=1$ transitions are mediated by quantum instanton
 tunneling at the rate $\exp (-4\pi/\alpha_W)$. However at high
temperatures the transitions can occur rapidly
via classical transitions induced by
thermal fluctuations over the now lower energy sphaleron.

The effective action for the gauge and Higgs fields at temperature $T$
contains the CP-violating Chern-Simons term \cite{TZ90}
\begin{equation}
\label{effCH}
\Delta S_{CP} \propto \int d^{4}x \,f(\phi, T) F^{\mu\nu}
{\tilde F}_{\mu\nu}
\end{equation}
where  ${\tilde F}^{\mu\nu}
 = \varepsilon ^{\mu\nu\rho\lambda}F_{\rho\lambda}$,
and $f(\phi, T)$ is a gauge invariant function constructed from the Higgs
doublets $\phi $ of the model. This term is heavily suppressed in the
standard model, being of order $10^{-20}$ \cite{SHAP8788}. However it
has been argued \cite{TZ90,MSTV91} that it may be much larger in
multi-Higgs doublet models with soft CP-violation in the Higgs sector.
Thus these extended electroweak models meet two of Sakharov's \cite{SAK}
conditions for generation of a matter--antimatter asymmetry, that is the
existence of (anomalous) $B$ violation and of $CP$ violation. The third
condition, the departure from thermal equilibrium, may occur during
the electroweak phase transition if it is first
order \cite{DHS92,Dine92,KRS85,P92}.

An interesting explicit scenario for the generation of a baryonic
asymmetry has been suggested \cite{TZ90,MSTV91,TZ91,TZ92} in which the
Chern-Simons term (\ref{effCH}), arising from soft CP-violating
terms in the Higgs potential, has the form of a CP-odd Higgs phase
linearly coupled to $N_{CS}$ and modifies the classical equations of
motion for the gauge fields such that, during the electroweak phase
transition, an asymmetry is produced in the probabilities of the
topology changing processes associated with $\Delta N_{CS} = +1$ and
$-1$, thereby driving the $N_{CS}$ value of the universe positive.

An alternative scenario has been advocated by Cohen {\em et al}
\cite{CK88,CKN91,CN92} in which the physical pseudoscalar field $\theta (x)$
orthogonal to the Goldstone boson, develops a derivative
coupling to the fermionic hypercharge current $j^{\mu}_{Y}$
\begin{equation}
\delta {\cal L} = 2 \partial_{\mu} \theta \, j^{\mu}_{Y}
\end{equation}
as a result of performing an anomaly-free space-time dependent rotation
on the fermion fields to remove $\theta (x)$ from the Yukawa couplings.
The spatial average of $\dot{\theta}$ acquires a non-zero
value during the electroweak phase transition and acts as a
potential for fermionic hypercharge which produces a free energy which
is minimised for non-zero baryon number. This mechanism requires
a first order phase transition, in order that the produced baryon
asymmetry be preserved against sphaleron transitions, which for
the standard model and MSSM translates into a stringent bound on the
physical Higgs masses ($m_H<55$ GeV \cite{DHS92}, and $m_H<64$ GeV
\cite{M92} respectively).

In general it is possible to only couple  $\dot\theta$ to hypercharge,
when only one Higgs is responsible for the fermion masses.
Because of this, the mechanism has not yet been
applied explicitly to supersymmetric models.
In this paper we examine the relation between these two effects.
First we extend the results of Ref.\cite{CKN91} to the most general case.
We then rederive the main
result of Ref.\cite{TZ90}, using the rotated-phase approach of
Ref.\cite{CKN91}, and find this effect to be negligible compared
with the former as anticipated in earlier work.
We conclude by applying the results to the minimal supersymmetric
standard model (MSSM) and the extended model with gauge singlet (ESSM).

\section{The two-Higgs model}

In order to clarify the connection between the approaches described
above,
we consider the most general soft CP-violating, two Higgs doublet
model \cite{GHKD90}. As in the supersymmetric Standard Model,
one doublet field, $\phi_{1}$, gives mass to the up-type quarks
and the other, $\phi_{2}$, to the down-type quarks and charged
leptons. The zero temperature Higgs potential is
\begin{eqnarray}
\label{Higgspot}
 V_{0} & = & \lambda_{1}(\phi_{1}^{\dagger}\phi_{1}-v_{1}^{2})^{2}
 + \lambda_{2}(\phi_{2}^{\dagger}\phi_{2}-v_{2}^{2})^{2} \nonumber \\
 &  &
  + \lambda_{3}(\phi_{1}^{\dagger}\phi_{1}-v_{1}^{2}+\phi_{2}^{\dagger}
\phi_{2}-v_{2}^{2})^{2} \nonumber \\
 &  &
 + \lambda_{4}[(\phi_{1}^{\dagger}\phi_{1})(\phi_{2}^{\dagger}\phi_{2})-
(\phi_{1}^{\dagger}\phi_{2})(\phi_{2}^{\dagger}\phi_{1})] \nonumber \\
 &  &
 + \lambda_{5}[\mbox{Re}
(\phi_{1}^{\dagger}\phi_{2})-v_{1}v_{2}\cos\xi]^{2}
\nonumber \\
 &  &
 + \lambda_{6}[\mbox{Im}
(\phi_{1}^{\dagger}\phi_{2})-v_{1}v_{2}\sin\xi]^{2}
\end{eqnarray}
With suitable values of the parameters $\lambda_{k}$ (all positive
for example), this potential is minimised for fields, up to
$SU(2)\times U(1)$ gauge transformations, of the form
\begin{equation}
\label{vacevs}
\langle \phi_{1} \rangle = \left(
\begin{array}{c}
            0 \nonumber \\
            \nonumber \\
            v_{1}
           \end{array}
\right)
\hspace*{2cm}
\langle \phi_{2} \rangle = \left(
\begin{array}{c}
          0 \nonumber \\
          \nonumber \\
          v_{2}e^{i\xi}
         \end{array}
\right)
\end{equation}
This potential softly breaks the reflection symmetries $\phi_{1} \rightarrow -
\phi_{1}$ and $\phi_{2} \rightarrow - \phi_{2}$. The
most general potential of this nature would also have a term
$\lambda_{7}\mbox{Im}
[(\phi_{1}^{\dagger}\phi_{2})^{2}]$ but this would induce
a non-zero upper component into one of the fields and thereby give rise
to an electrically charged vacuum.

At finite temperature and in an electrically neutral plasma we do not
anticipate any charge inducing terms in the free energy density
(temperature dependent potential) so that a minimum free energy field
configuration should also be of the form (\ref{vacevs}).

Electrically charged fields are not of interest to us in this paper and,
as we wish to consider neutral fields not only at a minimum of the free
energy but also during the phase transition connecting the minimum free
energies of the unbroken and broken phases, we take the neutral fields
to have the form
\begin{equation}
\phi_{1} = \left( \begin{array}{c}
0 \nonumber \\ \nonumber \\ \rho_{1}e^{-i\theta_1}
                  \end{array}
\right)
\hspace*{2cm}
\phi_{2} = \left( \begin{array}{c}
0 \nonumber \\ \nonumber \\ \rho_{2}e^{i\theta_2}
                  \end{array}
\right)
\end{equation}

The Lagrangian density describing the coupling of these Higgs fields
with the fermion fields is (using the two component Weyl notation)
\begin{eqnarray}
\label{Yukawa}
 {\cal L}_{m} & = & -\lambda_{u}\rho_{1}[u_{r}^{\dagger}e^{-i\theta_1}u_{l} +
u_{l}^{\dagger}e^{i\theta_1}u_{r}] \nonumber \\
&  & - \lambda_{d}\rho_{2}[d_{r}^{\dagger}e^{-i\theta_2}d_{l} +
d_{l}^{\dagger}e^{i\theta_2}d_{r}] \nonumber \\
&  & - \lambda_{E}\rho_{2}[E_{r}^{\dagger}e^{-i\theta_2}E_{l} +
E_{l}^{\dagger}e^{i\theta_2}E_{r}],
\end{eqnarray}
where a summation over generations (and consequently the KM matrix)
is implied.

Because of the presence of the interaction term
Im$(\phi_1^\dagger\phi_2)$ in Eq.(\ref{Higgspot}), away from the false
vacuum the Lagrangian is only invariant under changes in the
the relative phase $\alpha = \theta_1 + \theta_2$ of the two Higgs
doublets, and therefore a small variation in this
phase will be shared between $\theta_1$ and $\theta_2$ according to
\begin{eqnarray}
\label{par}
d\theta_1 = \frac{\rho_{2}^{2}}{\rho_{1}^{2}+\rho_{2}^{2}} d\alpha
\hspace*{2cm} d \theta_2 = \frac{\rho_{1}^{2}}{\rho_{1}^{2}+\rho_{2}^{2}}
d\alpha.
\end{eqnarray}
This mode of variation can be shown to be orthogonal to the Goldstone
mode which is absorbed into the gauge fields. Thus the partition
(\ref{par}) is the way that change in the phases should be
divided even when the fields are not at a free energy minimum, since
it is the only remaining physical degree of freedom once the
symmetry is broken. In terms of Lorentz covariants we have
\begin{eqnarray}
\label{defthetas}
\partial_{\mu}\theta_1 = \frac{\rho_{2}^{2}}{\rho_{1}^{2}+\rho_{2}^{2}}
 \partial_{\mu}\alpha \hspace*{2cm} \partial_{\mu}\theta_2 =
\frac{\rho_{1}^{2}}{\rho_{1}^{2}+\rho_{2}^{2}} \partial_{\mu}\alpha.
\end{eqnarray}

Following Cohen {\em et al} \cite{CKN91}, we would like
to absorb the phase factors
into a redefinition of the fermion fields.
In contrast with that work however, we do not have the option
of allocating them according to the fermion hypercharge, so that
we cannot avoid generating Chern-Simons terms involving the gauge
fields. In fact, the relative phases are
completely determined by the Higgs potential.
Therefore the only remaining freedom is the distribution of the
phase absorption between the
left and right handed fields. In order to allow for this,
we denote a generic Dirac field by
$\psi$. The most general rotation
\begin{equation}
\psi\rightarrow e^{i(a+b\gamma_5)\Theta} \psi,
\end{equation}
where $a$ and $b$ are constants,
generates the additional action term,
\begin{equation}
\label{ds0}
\delta S_0
 = -\int {\rm d}^4 x \left[
{\overline \psi}  m (e^{2 i  \Theta b\gamma_5}-1)
\psi+{\overline \psi} \gamma^\mu (a+b \gamma_5)
 \psi \partial_\mu \Theta \right].
\end{equation}
The axial rotation in the first piece will eventually cancel the
complex phase induced by the Higgs potential. The second piece
exhibits the conservation or otherwise of the associated current
via the Chern-Simons terms.
There are two Chern-Simons contributions to the
effective potential induced by the above rotation.
The first type is the anomaly term
(the $SU(2)$ pieces coming from the left handed rotations only) which
takes the form
\begin{equation}
\label{ds1}
\delta S_1
= \int {\rm d}^4 x \left[ i \Theta (a-b) \frac{F_l \tilde F_l}{32\pi^2}
                   -i \Theta (a+b) \frac{F_r \tilde F_r}{32\pi^2} \right]
\end{equation}
where $F_r$ $(F_l)$ is the gauge field coupling to $\psi_r$
$(\psi_l)$.
(The coupling constants have been absorbed into $F$ for
notational convenience, so that we may include semi-simple Lie
groups.)
Such terms may be calculated most efficiently using the invariant
path integral method as in \cite{F79}, and are temperature
independent as shown in Ref.\cite{RD85}.

In addition to this, the fermion loop diagram in Fig.(1) makes
a non-local contribution.
In order to determine the effective potential,
one makes a Taylor expansion in powers of momenta. This piece
then generates a Chern-Simons like term.
In calculating this contribution, our approach is
analagous to that used in Ref.\cite{TZ90}.
That is we consider the fields to be varying slowly enough with time
that we may assume a quasi-static equilibrium. This allows us to use the
techniques of finite temperature field theory to determine
the anomalous contribution to the free energy
density. In the calculation of Ref.\cite{TZ90}, the {\em time}
dependence of the field strengths was
extracted using the zeroth, ({\em temperature}) component of
momentum.
Therefore the finite temperature anomaly was obtained by making an
indirect appeal to the Lorentz invariance of the zero-temperature
object from which it came.
This procedure is outside the domain of standard finite temperature
field theory, and to avoid it, we instead examine the one loop
fermion diagrams shown in Fig.(1), and derive the $\Delta$
of Eq.(\ref{BL}) using the space components of the momenta only (for
technical details see for example Ref.\cite{kapusta}).
The contribution to the action may be expressed as
\begin{equation}
\label{ds2}
\delta S_2(T)
= \int {\rm d}^4 x \left[
i b\Theta\Lambda(T)\left(\frac{F_r \tilde F_r}{32\pi^2}
                   +\frac{F_l \tilde F_l}{32\pi^2}\right)\right],
\end{equation}
where
\begin{equation}
\Lambda(T)= \sum \frac{8\pi T m^2}{3(m^2+(2 n+1)^2 \pi^2 T^2)^{3/2}},
\end{equation}
and the sum runs over all integer
values of $n$. (We remind the reader that the expression above uses
the summation over space-time indices merely as a notational convenience,
and that finite temperature field theory has no Lorentz invariance.)
As pointed out in Ref.\cite{TZ90,TZ91}, there is a complementary
topological term generated in the Higgs sector, so that the total
topological number $N_{CS}-N_{H}$ is completely gauge invariant.
Such terms are generated by the two-loop diagram shown in Fig.(2).
At high temperature $T>>m$, we may make the approximation
\begin{equation}
\label{lambda}
\Lambda=\frac{14}{3}\zeta(3)
\frac{m^2}{\pi^2 T^2}
\left( 1-
\frac{93 \zeta(5)}{56 \zeta(3)}\frac{m^2}{\pi^2 T^2}
+\ldots \right).
\end{equation}
At this point we would like to add a cautionary note regarding
thermal calculations of this type.
On taking the zero-temperature limit $T\rightarrow 0$, we replace
\begin{equation}
\label{limit}
\sum f((2n+1)\pi T)\rightarrow \frac{1}{2\pi T}
\int {\rm d}\omega f(\omega)
\end{equation}
and recover the zero-temperature contribution
\begin{equation}
\label{zerotemp}
\delta S_2(0)
= \int {\rm d}^4 x \left[ i b\Theta \left(\frac{F_r \tilde F_r}{24\pi^2}
                   +\frac{F_l \tilde F_l}{24\pi^2}\right)\right],
\end{equation}
which may be recognized as the topological term in Ref.\cite{TZ91}.
Alternatively, one may take the zero-mass limit first, upon
which we find that $\delta S_2(T)=0$ for all temperatures, which
seems to contradict Eq.(\ref{zerotemp}). The resolution lies in the
fact that in order to make the calculation, we have implicitly
assumed that the mass interaction (i.e. the chirality--flip) remains
in equilibrium. When taking the $m\rightarrow 0$ limit, at some point
this will no longer be the case since (\ref{limit}) is only true
for $T<<m$. Thus Eq.(\ref{zerotemp}) will no longer be
valid.

On examining the total change in the action,
$\delta S=\delta S_0+\delta S_1+\delta S_2(T)$ we find that the
charge ($Q$) and baryon--lepton number ($B-L$) are conserved at
all temperatures,
but that  $Y$ and $B+L$ are not. Explicitly,
\begin{eqnarray}
\label{currents}
\partial_\mu j^\mu_{Q}   & = & 0\nonumber \\
\partial_\mu j^\mu_{B-L} & = & 0\nonumber \\
\partial_\mu j^\mu_{Y} +\ldots & = &
(\frac{3}{4} \Lambda_b-\frac{3}{4}\Lambda_t+\frac{1}{4}\Lambda_\tau)
\frac{i g^2_2 F_2 {\tilde F}_2}{32 \pi^2}
+(\frac{5}{48} \Lambda_b-\frac{17}{48}\Lambda_t+\frac{15}{48}\Lambda_\tau)
\frac{i g^2_Y F_Y {\tilde F}_Y}{32 \pi^2}
 \nonumber \\
\partial_\mu j^\mu_{B+L} & = &
4 N_g  \frac{i g^2_2 F_2 {\tilde F}_2}{32 \pi^2}
-N_g \frac{i g^2_Y F_Y {\tilde F}_Y}{32 \pi^2},
\end{eqnarray}
where the ellipsis stands for the mass terms in Eq.(\ref{ds0}),
 $F_2$ and $F_Y$ are the conventional $SU(2)$ and hypercharge
field strengths, and $N_g$ is the number of fermion generations.
Clearly the topological terms in the hypercharge equation
would vanish if we were able to take the
$T\rightarrow \infty$ limit. Physically what happens is that, as
the temperature is increased, the mean free path of particles in the
gas becomes much shorter than the path length of the (hypercharge
violating) chirality--flip. This is related to our previous point
concerning the zero mass limit of $\delta S_2$.

We are now ready to apply the method of Ref.\cite{CKN91} to the case
under discussion. First we remove the complex phases on the mass
terms in Eq.(\ref{Yukawa}) by making the rotations,
\begin{eqnarray}
u_r \rightarrow e^{i(\omega-\theta_1)}u_r &;&
u_l \rightarrow e^{i\omega}u_l\nonumber\\
d_r \rightarrow e^{i(\omega-\theta_2)}d_r &;&
d_l \rightarrow e^{i\omega}d_l\nonumber\\
E_r \rightarrow e^{i({\overline\omega}-\theta_2)}E_r &;&
E_l \rightarrow e^{i{\overline\omega}}E_l,
\end{eqnarray}
where we have introduced two arbitrary phases, $\omega$ and $\overline
\omega$. The corresponding changes in the action are,
\begin{eqnarray}
\label{ds00}
\delta S_0 & = & -\int {\rm d}^4 x \left[
{\overline u}_l \gamma^\mu u_l\partial_\mu \omega +
{\overline u}_r \gamma^\mu u_r\partial_\mu (\omega-\theta_1)+
{\overline d}_l \gamma^\mu d_l\partial_\mu \omega +
{\overline d}_r \gamma^\mu d_r\partial_\mu (\omega-\theta_2)
\right.
\nonumber\\
& & \left. +
{\overline E}_l \gamma^\mu E_l\partial_\mu {\overline\omega}+
{\overline E}_r \gamma^\mu E_r\partial_\mu ({\overline\omega}-\theta_2)+
{\overline \nu}_l \gamma^\mu \nu_l\partial_\mu {\overline\omega}+\ldots\right].
\end{eqnarray}
Again the ellipsis refers to the mass terms in Eq.(\ref{ds0}). In addition,
setting $\Lambda_b=\Lambda_\tau=0$, we have
\begin{equation}
\label{ds02}
\delta S_1+\delta S_2  =
\int {\rm d}^4 x
(18 \omega + 6{\overline\omega} - \frac{3}{2}\Lambda_t \theta_1)
\frac{i g_2^2 F_2 {\tilde F}_2}{32 \pi^2} -
(\frac{9}{2} \omega + \frac{3}{2}{\overline\omega}
- (4-\frac{17}{24}\Lambda_t) \theta_1-4 \theta_2)
\frac{i g_Y^2 F_Y {\tilde F}_Y}{32 \pi^2} .
\end{equation}
Instead of solving the equations of motion as in Ref.\cite{TZ90,TZ91},
we may remove the topological piece (i.e. the $SU(2)$ piece) by
making a judicious choice of $\omega$ in terms of $\overline \omega$, viz.
\begin{equation}
\omega= (\Lambda_t\theta_1-4{\overline\omega})/12.
\end{equation}
Notice that there is no solution which removes all contributions to
Eq.(\ref{ds02}), and there remains a term involving the hypercharge fields
which is given by,
\begin{equation}
\delta S_1+\delta S_2  =
\int {\rm d}^4 x
(4\theta_2+(4 - \frac{13}{12}\Lambda_t) \theta_1)
\frac{i g_Y^2 F_Y {\tilde F}_Y}{32 \pi^2} .
\end{equation}
In the above ,the angles $\omega$ and $\overline \omega$ are acting as
extra, space dependent $B$ and $L$ rotations. Thus this operation
may be interpreted as an absorption of the phase into the right
handed fields, followed by the removal of the topological $SU(2)$
piece by an appropriate $B+L$ rotation as given by Eq.(\ref{currents}).

Our analysis continues exactly as in Ref.\cite{CKN91}. First we substitute
$\omega$ into the current contributions given by (\ref{ds00}).
We then recalculate the net particle density imposing the above
constraints of conserved $B-L$ and $Q$, by including the chemical
potentials $\mu_{B-L}$ and $\mu_Q$. The relevant expression for
the density of a species $i$ is,
\begin{equation}
\rho_i=k_i (\langle\dot\Theta_i\rangle + (B_i-L_i)\mu_{B-L}
+Q_i \mu_Q) \frac{T^2}{6},
\end{equation}
where $\Theta_i$ is a generic symbol representing
the phases appearing in Eq.(\ref{ds00}).
The factor $k_i$ counts
colour and spin degrees of freedom, and factors of two for bosons
with respect to fermions.
This is a relativistic expansion, and there is no
Boltzmann suppression, since we are assuming that all species are
lighter than $T$ during the phase transition.
We now determine $\mu_Q$ and $\mu_{B-L}$
by solving $\rho_Q = \rho _{B-L} =0$, initially making the naive
assumption that all species and sphaleron transitions remain in equilibrium
during the phase transition;
\begin{eqnarray}
\mu_{B-L} & = & -\dot{\overline \omega}-\left(
(9-9\Lambda_t+3 n -\Lambda_t n)\langle \dot\theta_1 \rangle
+24\langle \dot\theta_2 \rangle)\right)/(111+13 n)\nonumber\\
\mu_{Q} & = & -\left(
(27+\frac{3}{4}\Lambda_t)\langle \dot\theta_1 \rangle
-\frac{39}{2}\langle \dot\theta_2 \rangle)\right)/(111+13 n)\nonumber\\
\rho_{B}=\rho_{L} & = & \frac{3}{2(111+13 n)}\left(
(162-51\Lambda_t+18 n -6\Lambda_t n)\langle \dot\theta_1 \rangle
+(210+26 n)\langle \dot\theta_2 \rangle)\right)
\frac{T^2}{6},
\end{eqnarray}
where $n$ is the number of light charged scalars. As expected, there
is no dependence of the final physical quantities, $\rho_B$ and
$\rho_{L}$, on the arbitrary angle $\overline\omega$. Thus the
mechanism does not depend on how the phase absorption is
divided between left and right handed fields.
As in Ref.\cite{CKN91}, non-zero phase-changes may bias the sphaleron
transitions to produce a net baryon number.

Clearly there are some potential inaccuracies involved in the
previous assumptions. The first arises from the fact that
the top is the only right handed particle expected to
remain in chemical equilibrium during the phase transition.
The fact that it is the left-handed particles which contribute to the
$SU(2)$ piece of the B+L non-conservation,
indicates that only the left handed fermions
are directly in chemical equilibrium with the sphalerons.
For right handed particles to be included, their masses
must be sufficiently large that they are able to perform
a chirality flip during the phase transition. We should also
take account of the fact that
there will not be time for the equilibrium densities to
be attained. Both these points are dealt with below,
where we introduce the relaxation equation which is governed by the
sphaleron transition rate.
In addition there is the inaccuracy due to the relativistic
expansion. Since the major contribution to the baryon
asymmetry occurs at the beginning of the phase transition (close to
the symmetric phase where the sphaleron transitions are unsuppressed)
where the physical top mass is small, this is an acceptable
approximation. Because of this it is probably more correct
to set $\Lambda_t\sim 0$ too.

\section{Comparing the two Approaches}

Our formulation of the Chern-Simons terms in the preceding section
allows an easy comparison of the method of Cohen {\em et al} applied
to this most general case, with that of Turok and
Zadrozny \cite{TZ90,TZ91}.
In order to give a more reasonable approximation
for the former, one should improve on the previous analysis by taking
into account the fact that the
top quark is the only right-handed particle in equilibrium during the
phase transition. Also we use the non-equilibrium rate equation
for the relaxation of the particle densities towards their equilibrium
values,
\begin{equation}
\dot\rho_B=-\Gamma_B \mu_B/T.
\end{equation}
$\Gamma_B$ is the baryon number violation per unit volume, and is given by
\begin{equation}
\Gamma_B=\kappa \alpha_W^4 T^4 {\mbox{e}}^{-4\pi V/g_2 T}.
\end{equation}
Setting the net densities of the light, right-handed particles to zero,
we find that
\begin{equation}
\mu_B= C(n) \langle {\dot\theta}_1\rangle
\end{equation}
where
\begin{equation}
C(n)=-\frac{3}{16}\frac{288-177\Lambda_t + 48 n -28 \Lambda_t n}
{159+25 n},
\end{equation}
Thus, making the step-function
approximation for $\Gamma_B$ as in Ref.\cite{CKN91}, we find that
\begin{equation}
\rho_B=C(n) \kappa \alpha_W^4 T^3 \Delta \theta_1,
\end{equation}
and therefore we can naturally have
\begin{equation}
\frac{\rho_B}{s}\sim 10^{-10}
\end{equation}
if $\lambda_5 \neq\lambda_6$, $\alpha\sim 10^{-2}$, and $\kappa \sim 1$.

In Ref.\cite{TZ90,TZ91}, sphaleron transitions
were biased by the one loop diagram contributions using a
linearised potential. Instead we may simply absorb the phases in
Eq.(\ref{Yukawa}) by rotating the right handed fields only, in order to
avoid creating $SU(2)$ anomalies (equivalent to setting
$\omega={\overline\omega}=0$ in Eq.(\ref{ds02})). Thus by
Eq.(\ref{ds2}) we find
\begin{equation}
\delta S_2 = -i \Delta
\int {\rm d}^4 x {\sqrt{\theta_1 \theta_2}}  F_2 {\tilde F}_2
\end{equation}
where
\begin{equation}
\label{del}
\Delta=\sum_{i}\frac{7 \zeta (3)}{8 \pi^2}
\left(\frac{M_W^2}{\pi^2 T^2}\right)
\lambda_i^2
\left( 1-93\zeta (5)/56\zeta(3)\frac{m_i^2}{\pi^2 T^2}+\ldots \right),
\end{equation}
where $M_W$ is the dynamical quantity, which changes during the
phase transition with $\rho_1$ and $\rho_2$.
This reproduces the result of Ref.\cite{TZ90},
and inserting the previous numerical values, we obtain
\begin{equation}
\frac{\rho_B}{s}< 2 \times 10^{-12}.
\end{equation}
Clearly, this contribution is suppressed by a factor of
$m^2_t/(4\pi^2 T^2)\sim10^{-2}$ which is the natural parameter of high
temperature expansions. Thus one should expect the
biasing in this case to be smaller than that caused by the
non-zero chemical potentials arising during the phase transition.
In addition one should bear in mind that $m_t$ (or
alternatively the $M_W$ appearing in Eq.(\ref{del})) is small
when the sphaleron transition rate $\Gamma_B$ is largest, so that
some further suppression may occur.
Since in both cases the final baryon asymmetry is proportional to
$\alpha$ (assuming that $\rho_1/\rho_2$ remains relatively constant during
the phase transition) it seems that the Turok and Zadrozny
effect should be less important as was anticipated in Ref.\cite{CKN91}.

\section{Supersymmetric Theories}

To conclude we examine the baryon asymmetry for the two supersymmetric
theories mentioned in Ref.\cite{CKN91}, namely the MSSM
and the ESSM.
In the first case the superpotential is of the form,
\begin{equation}
\label{mssmpot}
 {\cal W}  =  \lambda_{1} Q^{\dagger}_L {\overline H}_1 U_R
+ \lambda_{2} Q^{\dagger}_L H_2 D_R + \lambda_3 L^{\dagger} H_2 E_R
+ \mu ({\overline H}_1^0 H_2^0 - {\overline H}_1^+ H_2^-),
\end{equation}
where capital letters implies superfields.
After including the soft supersymmetry breaking terms we find that
the tree-level neutral Higgs potential is of the form,
\begin{eqnarray}
V_0  & = &  (g_2^2 + g_Y^2) (\rho_1^2-\rho_2^2)^2 /8 + |\mu|^2 (\rho_1^2 +
\rho_2^2)\nonumber \\
& & + m_1^2 \rho_1^2 +m_2^2 \rho_2^2 + 2 |B\mu| \rho_1 \rho_2 \cos
(\theta_2 - \theta_1 + \theta_{\mu} + \theta_{B}).
\end{eqnarray}
Throughout the phase transition, the minimum is given by
\begin{equation}
\theta_2 - \theta_1 = \pi-\theta_{\mu} - \theta_{B}
\end{equation}
and therefore for non-trivial values of $\rho_1$ and $\rho_2$, we must
have $\dot{\theta}_1=\dot{\theta}_2=0$ and so no spontaneous
baryogenesis.
Alternatively, as in Ref.\cite{DGH85}, we could have made a rotation of the
Higgs fields to remove the phase on $B\mu $ (that is $ \theta_{\mu}
+ \theta_{B} = 0$), without affecting the Kobayashi--Maskawa (KM) mixing
matrix. We expect that a more full analysis, along the lines of
Ref.\cite{CN92} for the anomaly free (hypercharge coupled) theory, would
yield a barely acceptable baryon asymmetry due to one loop corrections to the
effective potential. As that work showed, the additional requirements of
a sufficiently first order phase transition, and a small neutron
electric dipole moment, severely constrain the model.

In the ESSM, the electroweak symmetry breaking is provided by the
extra gauge singlet superfield $\Phi$. The superpotential is
\begin{equation}
\label{essmpot}
 {\cal W}  =  \lambda_{1} Q^{\dagger}_L {\overline H}_1 U_R
+ \lambda_{2} Q^{\dagger}_L H_2 D_R
+ \lambda_3 L^{\dagger} H_2 E_R
+ \lambda_7 \Phi ({\overline H}_1^0 H_2^0
- {\overline H}_1^+ H_2^-) + \lambda_8 \Phi^3.
\end{equation}
Letting the scalar component of the gauge singlet be $\Phi|=\rho_x e^{i
\theta_x}$, we find a neutral scalar Higgs potential of
\begin{eqnarray}
\label{vpot}
V_0  & = &  (g_2^2 + g_Y^2) (\rho_1^2-\rho_2^2)^2 /8 +
|\lambda_7|^2 \rho_x^2 (\rho_1^2 + \rho_2^2)+ 9 |\lambda_8|^2 \rho_x^4
+|\lambda_7|^2 \rho_1^2 \rho_2^2
\nonumber \\
& & + m_x^2 \rho_x^2 + m_1^2 \rho_1^2 +m_2^2 \rho_2^2
+ 6 |\lambda_7 \lambda_8| \rho_1\rho_2\rho_x^2
\cos (\theta_2 - \theta_1 -2 \theta_x+\theta_{\lambda_7}- \theta_{\lambda_8})
\nonumber\\
& & +  2 |A_7 \lambda_7 | \rho_x \rho_1 \rho_2
\cos (\theta_2 - \theta_1 + \theta_x+\theta_{\lambda_7} + \theta_{A_7})
+  2 |A_8 \lambda_8 | \rho_x^3
\cos (3 \theta_x+\theta_{\lambda_8} + \theta_{A_8}).
\end{eqnarray}
We first use the two Higgs rotations to eliminate the phases on the
$\lambda_7$ and $\lambda_8$ couplings. The only physical phases remaining
are those on $A_7$ and $A_8$, the $\theta$ vacuum, and the super-KM and KM
phases \cite{DGH85}.
Minimising this potential with respect to phases gives the conditions
\begin{eqnarray}
0 & = &\sin(\theta_2 - \theta_1 + \theta_x + \theta_{A_7})
+ 3 \left| \frac{\lambda_8}{A_7} \right| \rho_x
\sin  (\theta_2 - \theta_1 -2 \theta_x) \nonumber \\
0 & = &\sin (3 \theta_x + \theta_{A_8})
- 3 \left| \frac{\lambda_7}{A_8} \right|
\frac{\rho_1\rho_2}{\rho_x} \sin  (\theta_2 - \theta_1 -2 \theta_x).
\end{eqnarray}
In the case that $\theta_{A_7}- \theta_{A_8}=\pi$ we have the
trivial minimisation (i.e. when all the cosines in Eq.(\ref{vpot}) are
$-1$) given by
\begin{eqnarray}
\theta_2-\theta_1 & = &\pi - \frac{2}{3}\theta_{A_7}\nonumber\\
\theta_x & = & - \frac{1}{3}\theta_{A_7},
\end{eqnarray}
in which case there is again no spontaneous baryogenesis, since the
$A_7$ coupling is approximately constant during the phase transition.
(We make this last assumption because the phase transition is taken to
be first order, in which case it proceeds by spinodal decomposition
implying that the temperature remains nearly constant.) Now let
$\theta_{A_7}- \theta_{A_8}=\pi + \Delta \theta_A$, where we assume that
$\Delta \theta_A$ is small. Then the minimum may be approximated by
\begin{eqnarray}
\theta_2-\theta_1 & = &\pi - \frac{2}{3}\theta_{A_7}
-\Delta\theta_A \frac{A_8 \rho_x ( 6 \lambda_8 \rho_x - A_7)}
{3(3 A_7 \lambda_7 \rho_1 \rho_2 + A_7 A_8 \rho_x + 3 A_8 \lambda_8 \rho_x^2)}
\nonumber\\
\theta_x & = & - \frac{1}{3}\theta_{A_7}
-\Delta\theta_A \frac{A_8 \rho_x ( 3 \lambda_8 \rho_x + A_7)}
{3(3 A_7 \lambda_7 \rho_1 \rho_2 + A_7 A_8 \rho_x + 3 A_8 \lambda_8 \rho_x^2)}.
\end{eqnarray}
It should be noted that the residual angle appearing in the mass terms
may be absorbed by fermion rotations without affecting the KM matrix,
exactly as in the MSSM.
If we again make the assumption that the vacuum expectation values are
small when the baryon number violation is greatest, we may approximate
$\theta_2-\theta_1=\pi- (2 \theta_{A_7}-\Delta\theta_A)/3$, again
implying no baryon asymmetry. Clearly the production of a net baryon
number depends on how long the sphaleron transitions are allowed to continue
away from the symmetric phase. In order to make an estimate, we shall
use the renormalisation results of Ref.\cite{ASW92}, for the supergravity
inspired model with the parameters (in the notation of that paper),
$m_{1/2}=200$ GeV, $m_{0}=300$ GeV, $\tan\beta=2$, $\tan\beta_x=1/2$.
In addition we use the values for the couplings
$\lambda_7=g_2$, $\lambda_8=0.2$. We may induce a non-zero
$\Delta\theta_A$ by allowing a small phase on the
trilinear (soft-supersymmetry breaking) parameter
$A=|A|\exp i\Delta \theta_{A_0}$. We shall choose $|A|=0.2$.

Close to the GUT scale
$\Delta\theta_{A}=0$, but as the renormalisation group
equations are run down to the weak scale, a non-zero phase
develops. At the beginning of the phase transition
since the phases on $A_7$ and
$A_8$ are constant, there is no baryon number production. Towards
the end of the phase transition however, the phases are changing rapidly
due to the large vacuum expectation values that the Higgs fields have
aquired. The total change in phases is
\begin{equation}
\Delta\theta_2 - \Delta\theta_1  = 3.3 \Delta\theta_{A_0}
\end{equation}
and hence,
\begin{equation}
\frac{\rho_B}{s}\sim 10^{-8} \Delta\theta_{A_0}.
\end{equation}
In this case, the observed baryon asymmetry may quite easily
be generated, provided that sphaleron transitions are not
frozen out too quickly. In addition, we note from
Ref.\cite{P92} that the phase transition may be naturally first
order, since there are trilinear terms in the Higgs potential
even at tree level. This also implies that there may be
some enhancement of the baryon asymmetry arising from transport
processes as described in Ref.\cite{CKN91}.

\vspace{1cm}
\noindent
{\bf\Large Acknowledgement} \hspace{0.3cm} We would like to thank L.~McLerran
for useful discussions, and S.~Sarkar for a critical reading of the
manuscript. Two of us [SAA,WNC] would like to thank James Cook University
for hospitality extended during the completion of this work.

\pagebreak

\newpage
\section*{Figures}

\begin{description}

\item{\bf Figure 1 } The triangle diagram leading to anomalous
Chern-Simons terms in the effective potential, and also to finite
temperature non-anomalous contributions.

\item{\bf Figure 2 } The Higgs equivalent to Fig.(1), which leads
to a Higgs topological term, $N_H$. This ensures that the quantity
$N_{CS}-N_H$ is total gauge invariant which may be described as
the total derivative of a density, the Goldstone-Wilczek density.

\end{description}

\end{document}